\documentclass[fleqn,twoside]{article}
\usepackage{espcrc2}

\usepackage{amssymb}
\usepackage{amsbsy}
\usepackage{amsfonts}

\usepackage{graphicx}
\newcommand{\AmS}{{\protect\the\textfont2
A\kern-.1667em\lower.5ex\hbox{M}\kern-.125emS}}

\newcommand{\be}{\begin{equation}}
\newcommand{\ee}{\end{equation}}
\newcommand{\bea}{\begin{eqnarray}}
\newcommand{\eea}{\end{eqnarray}}

\newcommand{\eqref}[1]{(\ref{#1})}

\title{
\vspace{-5mm}
\rightline{\small MPP-2003-76}
The long range properties of the compact U(1) lattice gauge theory 
with the multi-level algorithm
}

\author{
Y. Koma\thanks{Talk presented by Y. Koma.
}, M. Koma and P. Majumdar
\address{Max-Planck-Institut f\"ur Physik,
F\"ohringer Ring 6, D-80805, M\"unchen, Germany}
}

\begin{document}

\begin{abstract}
    The 4D compact U(1) lattice gauge theory (LGT)
    in the confinement phase is studied
    with the multi-level algorithm.
    The static potential, force and flux-tube profile    
    between two static charges
    are precisely measured from correlation functions
    involving the Polyakov loop.
    Universality of the coefficient of the $1/r$ 
correction to the static potential, known as the 
L\"uscher term, and the transversal width of the flux-tube 
profile are investigated.
\end{abstract}

\maketitle

\section{Introduction}
\vspace*{-0.1cm}

\par
If the confining gauge theory is related to an 
effective bosonic string
theory, the long range behavior of the potential
between static charges separated by distance $r$
is expected to have the form:
\be
V(r) = \sigma r + \mu + \frac{\gamma}{r}+O(\frac{1}{r^{2}}) .
\label{eqn:asympt-pot}
\ee
Here $\sigma$ is the string tension and
$\mu$ denotes a constant.
The third term is known as the L\"uscher term~\cite{Luscher:1980fr}, 
where the coefficient, $\gamma = -\pi (d-2)/24$ 
with space-time dimension $d$,
is considered to be universal.
The effective bosonic string theory
also predicts that the width of the field energy distribution
of the flux tube diverges logarithmically as 
$r \to \infty$~\cite{Luscher:1981iy}.
Recent Monte-Carlo simulations of various
LGTs  support Eq.~\eqref{eqn:asympt-pot} 
with high accuracy: 
the confinement phase of
$Z\!\!\!Z_{2}$ LGT~\cite{Caselle:2002ah},
SU(2) LGT~\cite{Bali:1995de} and
SU(3) LGT~\cite{Juge:2002br}.

\par
In this report we investigate the long range properties of
compact U(1) LGT in the confinement phase.
From the Polyakov loop correlation function (PLCF),
we extract the potential and force to see whether this theory
also supports the presence of the 
universal correction to the static potential.
We then measure the profile of the flux tube
induced by the PLCF, and
see the behavior of its width as a function of $r$.
However, due to its strong coupling nature, the Monte Carlo
simulation of compact U(1) LTG is numerically difficult,
when its long range properties are of interest.
To obtain reliable signals, we apply here the
multi-level algorithm proposed by
L\"uscher and Weisz originally for SU(3)
LGT~\cite{Luscher:2001up},
which helps to reduce statistical errors
exponentially.

\section{Numerical procedures}
 
We adopt the terminology of Ref.~\cite{Luscher:2001up}.
To measure the PLCF, $\langle P^{*} P (R) \rangle$,
where $R = r/a$, we take 
second-level averages of the two-link correlators
$\mathbb{T}(m;R;i)= U_{4}^{*}(m) U_{4}(m+ R \; \hat{i})$
with $m=(m_s,m_t)$
as
\be
\mathbb{T}^{(2)}(m_{s},\bar{m}_{t}; R;i)
= [\mathbb{T}(m ; R;i)\mathbb{T}(m+\hat{4};R;i) ],
\label{eq:two-link2}
\ee
which is achieved by updating link variables 
except for the spatial links 
at $\bar{m}_{t}=1,3,\ldots, N_{t}-1$.
$N_{t}$ is the timelike extent of the lattice volume.
$i=1,2,3$ denote directions between the two charges.
We call this procedure the {\em internal update}.
We repeat the internal update until  
reasonably stable values for $\mathbb{T}^{(2)}$ are 
obtained.
Then the PLCF at a spatial site $m_{s}$ is constructed as
\bea
&&
\!\!\!\!\!\!\!\!\!\!\!\!
P^{*} P (m_{s};R;i) \! = \!
{\rm Re} \; 
\mathbb{T}^{(2)}(m_{s},1;R;i) \nonumber\\
&&
\!\!\!\!\!\!\!\!
\times
\mathbb{T}^{(2)}(m_{s},3 ;R;i)
\cdots
\mathbb{T}^{(2)}(m_{s}, N_{t} - 1;R;i).
\label{eqn:poly1}
\eea
The average with respect
to $m_{s}$ and $i$ provides
$[ P^{*} P (R) ]_{i_{c}}$.
The desired expectation value 
is the average of $[ P^{*} P (R) ]_{i_{c}}$
for $i_{c}=1,2,\ldots,N_{c}$.
In the actual measurements, 
we have also applied the multi-hit 
technique to the timelike link variables for $R \ge 2$ before
constructing  $\mathbb{T}$.
The static potential and the corresponding force 
are taken as (neglecting terms of 
$O(e^{-\Delta E N_{t}})$)
\bea
&&\!\!\!\!\!\!\!\!\!\!\!\!
aV(R) = -\frac{1}{N_{t}} \ln 
\langle P^{*}P (R) \rangle \; ,
\label{eqn:potential}
\\
&&\!\!\!\!\!\!\!\!\!\!\!\!
a^{2} F(\bar{R}) = aV(R)-aV(R-1) 
\; ,
\label{eqn:force}
\eea
where $\bar{R}=R-1/2 + O(a)$.

\par
In order to measure the flux-tube profile, one needs
to compute a correlation function of the type
\bea
\langle  {\cal O}(n)  \rangle_{j} =
\frac{\langle  P^{*}P {\cal O}(n) \rangle_{0}}
{\langle P^{*}P \rangle_{0}} - \langle {\cal O}\rangle_{0} \; ,
\label{eq:profile-correlator}
\eea
where ${\cal O}$ is a local operator,
$\langle \cdots \rangle_{j}$ denotes an average 
in the vacuum with the PLCF,
and $\langle \cdots \rangle_{0}$ an average in the vacuum 
without such a source.
To measure $\langle  P^{*}P {\cal O} \rangle_{0}$
on the mid-plane between two static charges, 
we parameterize the position of the local operator $n$ 
as
$n= (
n_{i}\! = \! m_{i} \! + \! R/2, \;
n_{j}\! = \! m_{j} \! + \! x, \;
n_{k}\! = \! m_{k}\! + \! y, \; 
n_{t}\! = \! m_{t}) $
and  take the average 
of the two-link-local-operator correlator
\bea
&&\!\!\!\!\!\!\!\!\!\!\!\!
\mathbb{TO}^{(2)}(m_{s},\bar{m}_{t} ; n ; R ; i) 
=
\nonumber\\&&\!\!\!\!
[\mathbb{T}(m; R ;i )\mathbb{O}(m+\hat{4}; n ; R; i) ]\; .
\label{eqn:ope-te}
\eea
with $\mathbb{O}(m; n ; R; i) 
=  U_{4}^{*}(m) U_{4}(m +R \hat{i}) {\cal O}(n)$
in addition to $\mathbb{T}^{(2)}$.
Combining $\mathbb{TO}^{(2)}$ and $\mathbb{T}^{(2)}$,
we obtain $\langle  P^{*}P {\cal O} \rangle_{0}$.

\par
As local operators, we have used 
${\cal O}_{ E}(n)= i \bar{\theta}_{\mu\nu}(n)$ for
the electric field, 
and ${\cal O}_{k}(n)= 2 \pi i k_{\mu}(n)$
for the monopole current
as defined in Ref.~\cite{DeGrand:1980eq}.
Note that the second term of Eq.~\eqref{eq:profile-correlator}
vanishes for these operators 
since $\mathcal{O}_{E}$ and $\mathcal{O}_{k}$ are
parity odd.

\section{Numerical results}

\par
We generate a sequence of independent gauge field 
configurations, by using
the Wilson gauge action of compact U(1) LGT on a $16^{4}$ 
lattice at $\beta =$  0.98, 0.99, 1.00, 1.005 and 1.01.
For the first three $\beta$ values,
the configuration has been generated
after 500 thermalization sweeps, 
and they were separated by 100 sweeps,
where one Monte Carlo update has been 
achieved by 1HB/3OR.
For $\beta=$ 1.005 and 1.01 
the thermalization sweeps have been 
taken 1000 and 3000, respectively,
with 1HB/5OR Monte Carlo update.

\par
For the static potential and force, 
we have used $N_{c}=$ 1050, 1250, 2050, 2400 and 3200 
configurations, respectively.
The internal updates 
have been taken as $N_{\rm iupd}=$10000, 8000, 
5000, 3000 and 1000.
These numbers are optimized to measure the PLCF
up to  $R$ = 6 with $< 10$ \% errors, where
the PLCF themselves
take values from $10^{-3}$ to $10^{-17}$.
From the force, we have introduced a scale 
based on Sommer's relation $r_{0}^{2}F(r_{0}) = 1.65$.
Lattice spacings in units of $r_0$ are found to be 
$a/r_0=$ 0.470, 0.422, 0.353, 0.305 and 0.217, respectively.

\begin{figure}[!t]
\includegraphics[height=5cm]{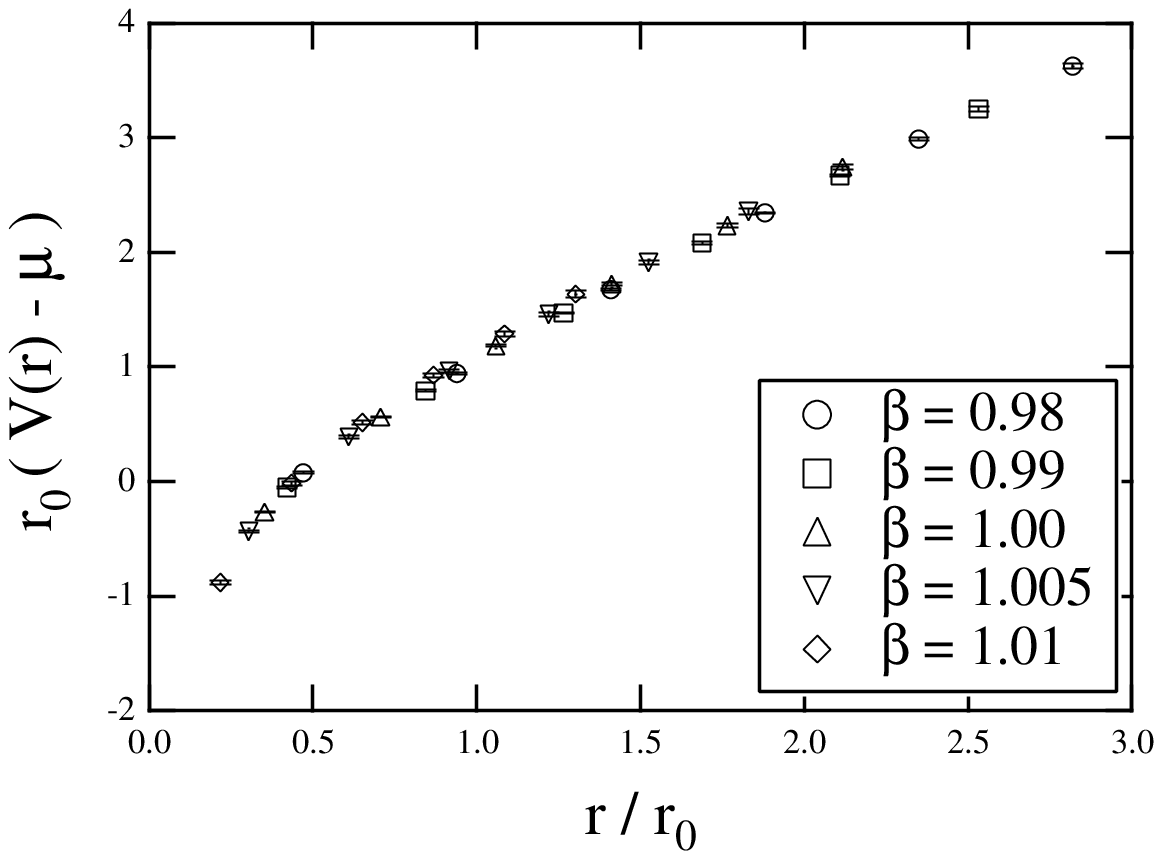}
\vspace*{-1cm}
\caption{The static potential as a function of $r/r_0$.
The constant $\mu$ is subtracted by using the fit result 
of Eq.~(1), neglecting $O(1/r^{2})$ corrections.}
\label{fig:pot}
\includegraphics[height=5cm]{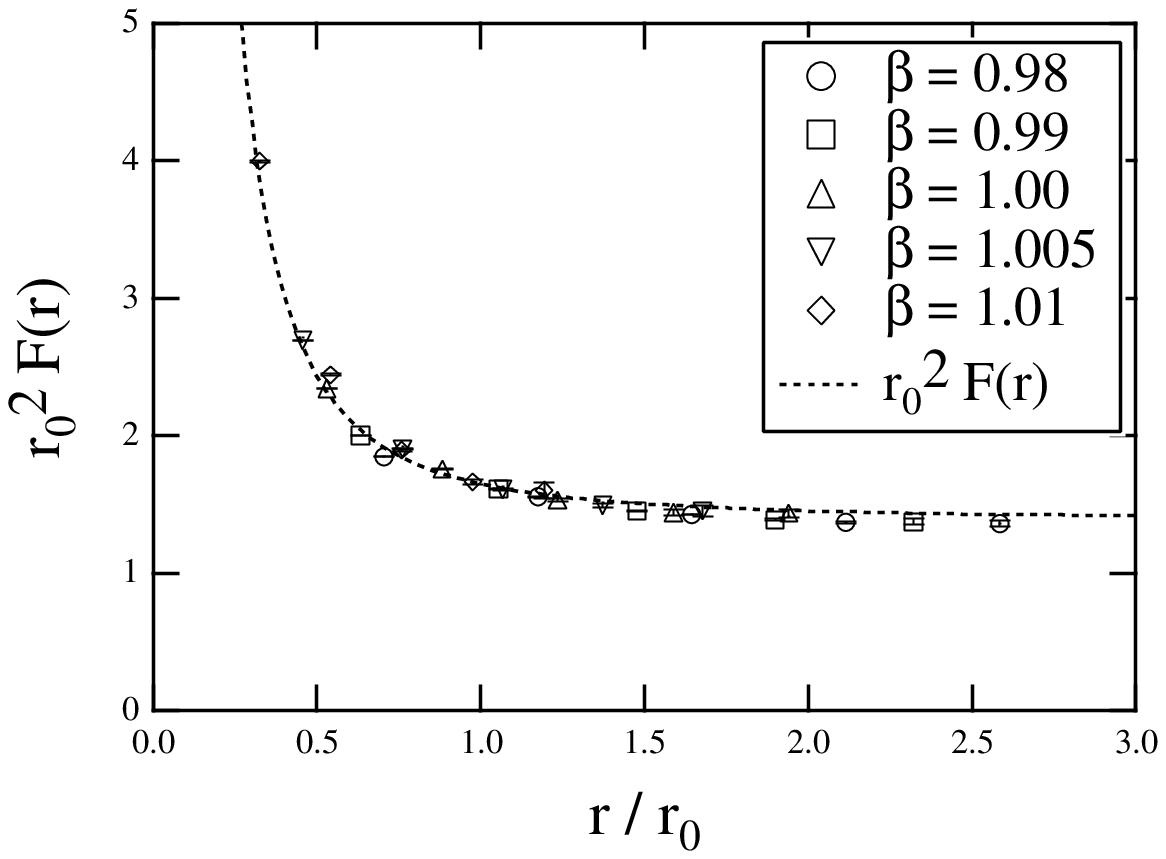}
\vspace*{-1cm}
\caption{The force as a function of $r/r_0$.
The dotted line corresponds to the theoretical 
prediction of Eq.~(1) up to $O(1/r^{2})$ corrections.}
\label{fig:force}
\vspace*{-0.5cm}
\end{figure}

\par
In Figs.~\ref{fig:pot} and \ref{fig:force},
we show the static potential and force from all $\beta$ values.
They show good scaling behaviors.
We also plot the theoretical prediction for the force
based on Eq.~\eqref{eqn:asympt-pot} in~Fig.~\ref{fig:force}.
We find that the long distance behavior is well-described by
the function, $F = dV/dr = \sigma - \gamma/r^{2}$ with
$\sigma r_{0}^{2} = 1.65 - \gamma = 1.39$, which
contains no fitting parameter.
This supports the universality of $\gamma/r$ correction 
to the static potential.
Surprisingly, another feature in common with non-Abelian gauge 
theories is that 
this function also fits the data down to relatively short 
distance to $r/r_{0} \sim 0.3$.
For this, there is as yet no explanation.

\par
\begin{figure}[!t]
\includegraphics[height=5.cm]{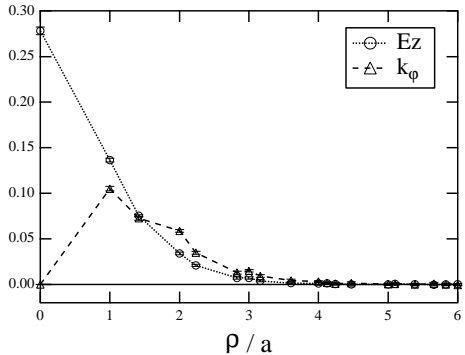}
\vspace*{-1.1cm}
\caption{The profiles of the 
electric field ($E_{z}$) and monopole current ($k_{\varphi}$) 
for  $R=5$ at $\beta = 0.98$ as a function of the radius of 
the flux tube $\rho$.} 
\label{fig:profile}
 \vspace*{-0.3cm}
\end{figure}

We have measured  flux-tube profiles with 
lengths $R=$ 3 to 6 at $\beta=$ 0.99 to 1.01 and 
$R = 3$ to 5 at $\beta = 0.98$.
We have applied $N_{\mathrm{iupd}}
= 200 \sim 8000$ depending on 
$R$ and $\beta$.
The number of configurations are $N_{c}=300$ for all $\beta$ values.
As an example, we show the profiles of the 
electric field  and monopole current 
for $R=5$ at $\beta=0.98$ in Fig.~\ref{fig:profile},
which is the longest flux tube in these 
measurements: $r/r_0 = 2.35$.
Here we have taken the cylindrical average: $\rho=\sqrt{x^{2}+y^{2}}$
and $\varphi = \tan^{-1} (y/x)$.
As shown in this figure, we have
obtained very clean signals, however, rotational invariance
for the monopole current profile
seems not good due to the relatively large lattice 
spacing at $\beta = 0.98$.
For larger $\beta$, we obtained  smoother curves.

\begin{figure}[!t]
\includegraphics[height=5.cm]{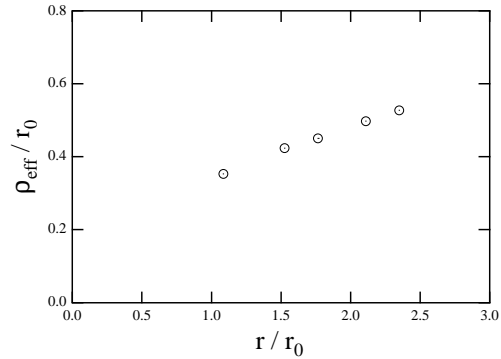}
\vspace*{-1.1cm}
\caption{The peak radius 
of the monopole current profile as a function of $r/r_{0}$.}
\label{fig:kwidth}
\vspace*{-0.3cm}
\end{figure}

There are several definitions of the width of the flux tube.
Here, we have investigated this in terms of 
the peak radius of the monopole current profile, $\rho_{\rm eff}$.
In order to find $\rho_{\rm eff}$ we have
interpolated the on-axis data with a smooth curve.
The result as a function of  $r/r_0$
is shown in Fig.~\ref{fig:kwidth}, where
we have picked only the data with $R=5$ for all  $\beta$ values; 
$\rho_{\rm eff}$ seems to  grow as increasing $r$.
However, we make no statement on the behavior 
whether the  width  diverges as $r \to \infty$.
Also, since the rotational invariance is not good 
for small $\beta$ values, we have to check the validity
of the smooth interpolation carefully.
Further detailed analyses are in progress~\cite{Koma}.

\subsection*{Acknowledgement}
We are grateful to M.~L\"uscher and P.~Weisz 
for useful discussions. 
The calculations were done on the 
SX-5 at the RCNP, Osaka University, Japan.

\end{document}